\documentclass[pra,showpacs,showkeys]{revtex4}
\usepackage{graphicx}
\usepackage{indentfirst}
\usepackage{dcolumn}
\usepackage{amsfonts}
\usepackage{amsmath}
\usepackage{bm}
\newcommand{\ket}[1]{\left|#1\right\rangle}
\newcommand{\bra}[1]{\left\langle#1\right|}
\newcommand{\bgeq}{\begin{equation}}
\newcommand{\edeq}{\end{equation}}
\newcommand{\bgeqn}{\begin{eqnarray}}
\newcommand{\edeqn}{\end{eqnarray}}

\begin{document}

\title{Weak measurements with a qubit meter}

\date{\today}

\author{Shengjun Wu$^{1,2}$
and Klaus M{\o}lmer$^{1}$}
\affiliation{$^1$Lundbeck Foundation Theoretical Center for Quantum
System Research, Department of Physics and Astronomy, University of
Aarhus, DK-8000 \AA rhus C, Denmark
\\
$^2$Hefei National Laboratory for Physical Sciences at Microscale,
University of Science and Technology of China, Hefei, Anhui 230026, China}

\begin{abstract}
We derive schemes to measure the so-called weak values of quantum system observables by coupling of the system to a qubit meter system. We highlight, in particular, the meaning of the imaginary part of the weak values, and show how it can be measured directly on equal footing with the real part of the weak value. We present compact expressions for the weak value of single qubit observables and of product observables on qubit pairs. Experimental studies of the results are suggested with cold trapped ions.
\end{abstract}


\maketitle

\section{Introduction}

Weak measurements were introduced by Aharonov, Albert and Vaidman \cite{AAV88} to challenge the view that we can only ascribe physical reality to the value of a physical observable in quantum mechanics, if it is actually measured. With the introduction of an arbitrarily weak coupling to a suitable meter quantum system, it is possible to repeat an experiment a large number of times, and thus obtain a very precise value for the average change of the meter state, and hereby the average value of the relevant system observable. In the limit of a very weak coupling such a measurement causes no disturbance of the original quantum system, and as suggested by Aharonov et al, one may thus proceed and measure other observables, including of course also non-commuting observables on the same quantum system.

A central result in the weak measurement formalism is the expression
\bgeq
\left\langle A \right\rangle _w = \frac{\bra{\psi_f} A \ket{\psi_i}}{\left\langle \psi_f | \psi_i \right\rangle}.
\label{eq:weak}
\edeq
for the weak value of a system observable $A$, assuming that the system is initially prepared in the state $\left| \psi_i \right\rangle$, and is finally post-selected (by a projective measurement) in the state $\left| \psi_f \right\rangle$. Note that if the initial or the final state is an eigenstate of the operator $A$, the weak value merely becomes the corresponding eigenvalue, but in the general case of arbitrary $\left| \psi_i \right\rangle$, $\left| \psi_f \right\rangle$, the meaning of the outcome is less clear.

The expression (\ref{eq:weak}), follows from an analysis of an experimental procedure,
where the system in state $\left| \psi_i \right\rangle$ is coupled to another quantum system (the meter),
prepared in state $|\phi\rangle$, by an interaction Hamiltonian $H_{int}=\hbar g(t)A\cdot X$, where $X$ is some meter observable. To model a short duration interaction at time $t_0$, we will assume $g(t)=g\delta(t-t_0)$, and thus obtain the  time evolved state after the interaction,
\bgeq
|\Psi\rangle = e^{-igA\cdot X} \left| \psi_i \right\rangle |\phi\rangle,
\edeq
and if we subsequently measure the quantum system to be in the state  $\left| \psi_f \right\rangle$, the meter system is left in the un-normalized state
\bgeqn
|\alpha\rangle &=& \bra{\psi_f} e^{-igA \cdot X} \left| \psi_i \right\rangle |\phi\rangle \nonumber \\
&\approx & \bra{\psi_f} I-igA \cdot X \left| \psi_i \right\rangle \left|\phi \right\rangle \nonumber \\
&=& \left\langle \psi_f | \psi_i \right\rangle (1-ig \langle A \rangle_w\cdot X)|\phi\rangle \nonumber \\
&\approx & \left\langle \psi_f | \psi_i \right\rangle e^{-ig\langle A\rangle_w\cdot X}|\phi\rangle.
\label{eq:derivation}
\edeqn

In the following we will assume that both $g$, $X$ and $A$ are dimensionless or are expressed as dimensionless quantities. We thus see that after the interaction and post-selection the meter has
effectively experienced the weak value of the observable $A$ as a classical coupling strength, and hence a final measurement on the (re-normalized) meter will reveal the value $\langle A\rangle_w$. Weak values have been studied extensively theoretically and experimentally in the past decades and they have a number of interesting and thought-provoking properties: the weak value is the value, that one is led to ascribe to an observable $A$ by counterfactual reasoning, if one hypothesizes about the value of $A$ during the undisturbed evolution of the system between its initial preparation in $|\psi_i\rangle$ and final detection in $|\psi_f\rangle$ \cite{Popescu,Molmer01,YYKI08}; the weak value may be far outside the range of the eigenspectrum of the operator $A$; even though $A$ is hermitian, the weak value may be complex.

Weak measurements have been implemented and demonstrated by, for example, the lateral displacement of coherent beams of light by birefringent materials \cite{RSH91,HK2008}, and two-photon entangling operations with post-selection  \cite{Pryde05}. Various optical and quantum optical systems  are currently subject to increased interest as weak measurements both address questions of fundamental importance, related to Bohm trajectories \cite{Wiseman07}, the measurement-disturbance relationship in interferometric experiments \cite{Mir07}, and macro-realism explored by Leggett-Garg inequalities \cite{WJ08,GABLOWP09}. Furthermore, weak measurements are proposed to hold the potential to achieve higher precision than conventional measurements \cite{Dixon2009} and to modify in a novel way the entanglement contents in bipartite quantum systems \cite{Menzies2008}.

It is the purpose of this paper to throw more light on weak values, and in particular to describe possible experimental methods which can extract the real and imaginary part of the weak value expression (1). In Sec. II, we briefly review a theoretical analysis by Richard Jozsa \cite{Jozsa2007}, showing how a free particle used as a meter coupled to the quantum system of interest can extract both the real and imaginary part of the weak value, and we present the same analysis for the case of a qubit meter system, which provides a more symmetric treatment of the weak value quadratures. In Sec. III, we provide compact analytical expressions for weak value measurements on a qubit system, and we discuss measurements of product operators on pairs of qubits and comment on their factorization properties. In Sec. IV, experimental investigations are proposed with trapped atomic ions. Sec. V. concludes the paper.

\section{Measuring a complex weak value}

\subsection{Measurement by coupling to a free-particle meter system}

Although the complex valued $\langle A\rangle_w$ applies as a parameter in the evolution of the meter system in (\ref{eq:derivation}), the read-out of the meter must consist in a measurement of the change of a physical observable of that system which must in turn yield a real number. In an attempt to understand the complex valued $\langle A\rangle_w$, Richard Jozsa \cite{Jozsa2007}, recently considered the special case of a coupling term $X$ being the generator of translations (the momentum operator $p$) of a free particle meter system with position coordinate $q$ and mass $m$. This study showed that the mean displacement of the meter contained directly the real part of $\langle A\rangle_w$, and in addition a (real) term proportional to the product of the imaginary part of $\langle A\rangle_w$ and the instantaneous rate of change of the position variance of the particle at the interaction time $t_0$ \cite{Jozsa2007},
\begin{equation}
\langle q\rangle_f=\langle q\rangle_i + g Re(\langle A\rangle_w)+g Im(\langle A\rangle_w)m \frac{d}{dt}Var(q)|_{t_0}.
\end{equation}
The last term vanishes if the meter wave-packet is a real wave function with no local probability current components and it is normally disregarded in discussions of weak value measurements. Jozsa also pointed out in \cite{Jozsa2007} that the imaginary part of the weak value is directly available through a measurement of the change of the meter-"momentum" operator,
\begin{equation}
\langle p\rangle_f=\langle p\rangle_i + 2g Im(\langle A\rangle_w ) Var(p)|_{t_0},
\end{equation}
The last term in this expression does not vanish for a real wave function, but it becomes very small if the meter wave function is very broad in position space and very narrow in momentum space which was, in fact, originally proposed as the true weak-measurement limit \cite{AAV88}. Here, however, the smallness of the coupling parameter $g$ ensures this limit, and the meter wave packet can be chosen with finite width in both position and momentum space, so that the real and the imaginary part of the weak value can be determined experimentally at will. See also ref. \cite{augusto} for a quantum phase space perspective on real value measurements. We note, that in the same way as anomalous real weak values occur outside the spectrum of the operator $A$, an imaginary part of the expectation value of a Hermitian operator is anomalous, and, indeed, an interaction Hamiltonian $H_{int}=g(t)A\cdot p$ is normally not expected to change the value of the Quantum Non-Demolition (QND) meter observable $p$. It is of course the pre- and post-selection by different states of the quantum system that change these conventional properties. The enumerator in the weak value expression (1) can be rewritten as the expectation value in the initial state $|\psi_i\rangle$ of the product operator $|\psi_i\rangle\langle\psi_f | \cdot A$, and it is of course well known in quantum mechanics that the mean value of operator products does not coincide with the product of their mean values, and that, e.g.,  the product of two non-commuting Hermitian operators, like position and momentum, is not Hermitian, and may hence have a complex expectation value.

\subsection{Measurement by coupling to a qubit meter system}

Suppose we introduce a spin-$1/2$ particle as the measuring device, which is in the initial
state represented by a unit length Bloch vector $\hat{m}$, $\ket{\Phi_i}\bra{\Phi_i} =\frac{1}{2} \left( I +  \hat{m} \cdot \overrightarrow{\sigma_2}\right)$. We assume an interaction between the quantum system of interest and the qubit meter system, given by the Hamiltonian
\bgeq
H_{int} = g \hbar \delta (t-t_0) A \cdot (\hat{n} \cdot \overrightarrow{\sigma_2}),
\edeq
where $\hat{n}$ is a unit vector and  $g\ll 1$ is a dimensionless positive number.

After the interaction and post-selection of the first particle, the state of the second qubit is given by
\bgeqn
\ket{\Phi_f}_2 &=& \bra{\psi_f} e^{-i g A \cdot (\hat{n} \cdot \overrightarrow{\sigma_2})}
\ket{\psi_i} \ket{\Phi_i} \nonumber \\
&\approx & \left\langle \psi_f | \psi_i \right\rangle (I- i g \left\langle A \right\rangle _w
\cdot (\hat{n} \cdot \overrightarrow{\sigma_2})) \ket{\Phi_i}_2 . \nonumber
\edeqn
After the interaction and post selection, we are free to measure any observable on the meter qubit, i.e., any operator of the form $\hat{q}\cdot \overrightarrow{\sigma}_2$, and conditioned
on the post-selection of the quantum system of interest, the expectation value of this operator is therefore given by (to first order in $g$)
\bgeqn
\frac{\bra{\Phi_f} \hat{q}\cdot \overrightarrow{\sigma_2} \ket{\Phi_f}}{\left\langle \Phi_f | \Phi_f \right\rangle}
&\approx & \bra{\Phi_i} \hat{q}\cdot \overrightarrow{\sigma_2} \ket{\Phi_i} +
g \bra{\Phi_i} i [\hat{n}\cdot \overrightarrow{\sigma_2}, \hat{q}\cdot \overrightarrow{\sigma_2}] \ket{\Phi_i} \cdot
Re \left\langle A \right\rangle _w \nonumber \\
& &+g \left\{ \bra{\Phi_i}\hat{n}\cdot \overrightarrow{\sigma_2} \hat{q}\cdot \overrightarrow{\sigma_2} +
\hat{q}\cdot \overrightarrow{\sigma_2} \hat{n}\cdot \overrightarrow{\sigma_2}  \ket{\Phi_i}
-2 \bra{\Phi_i} \hat{n}\cdot \overrightarrow{\sigma_2} \ket{\Phi_i} \bra{\Phi_i} \hat{q}\cdot \overrightarrow{\sigma_2} \ket{\Phi_i}\right\} \cdot Im \left\langle A \right\rangle _w \nonumber \\
&=& \hat{q} \cdot \hat{m} + 2 g \{ (\hat{q} \times \hat{n}) \cdot \hat{m} \} Re \left\langle A\right\rangle _w + 2 g \{ \hat{n} \cdot \hat{q} - (\hat{n}\cdot \hat{m})(\hat{q} \cdot \hat{m}) \}  Im \left\langle A \right\rangle _w
\edeqn

We observe that by merely choosing the appropriate initial state of the meter qubit, the real part and imaginary part of the weak value $\left\langle A \right\rangle _w$
can be obtained as the measured expectation value of suitable spin projection operators on the meter qubit.
If we choose an initial state with a mean spin perpendicular to the axis of precession due to $H_{int}$, $\hat{m} \perp \hat{n}$, and detection of the spin component emerging due to this spin precession, $\hat{q} =\hat{n} \times \hat{m}$, then we can measure the real part of the weak value as a conventional, renormalized expectation value
\bgeq
2 g Re \left\langle A \right\rangle _w
\approx
\frac{\bra{\Phi_f} \hat{q}\cdot \overrightarrow{\sigma_2} \ket{\Phi_f}}{\left\langle \Phi_f | \Phi_f \right\rangle}
.
\label{eq:real1}
\edeq
If, on the other hand we choose the same initial state $\hat{m} \perp \hat{n}$ and a measurement of the spin projection along the axis of precession, $\hat{q} =\hat{n}$, then
\bgeq
2 g Im \left\langle A \right\rangle _w
\approx
\frac{\bra{\Phi_f} \hat{q}\cdot \overrightarrow{\sigma_2} \ket{\Phi_f}}{\left\langle \Phi_f | \Phi_f \right\rangle}
.
\label{eq:imag1}
\edeq

We note that similarly to the analysis in the previous subsection, the imaginary part of the weak value causes an "anomalous" change in the QND interaction variable, which is here the spin component $(\hat{n}\cdot \overrightarrow{\sigma_2})$ along the precession axis.

The expressions, (\ref{eq:real1},\ref{eq:imag1}) look very similar, but they actually reflect two very different processes occurring to the meter qubit.

The real part of the weak value is read out due to the action of the operator
\bgeq
e^{- i g Re \left\langle A \right\rangle _w
\cdot (\hat{n} \cdot \overrightarrow{\sigma_2})} \nonumber
\edeq
on the meter qubit, i.e., a real rotation around the $\hat{n}$-axis, which is read out by preparing the spin along a given direction and measuring the growth of its component in the orthogonal direction in the plane perpendicular to the axis of precession.

The imaginary part, however, is linked with the action of the operator
\bgeq
e^{g Im \left\langle A \right\rangle _w
\cdot (\hat{n} \cdot \overrightarrow{\sigma_2})}. \nonumber
\edeq
This is not a rotation and not even a unitary operator. The operator
$(\hat{n} \cdot \overrightarrow{\sigma_2})$ has two eigenstates with eigenvalues $\pm 1$,
and with an infinitesimal value of $g$, the action on the meter qubit is to increase/decrease
the amplitude of these eigenstates according to the value of
$Im \left\langle A \right\rangle _w$.
For an initial state along $\hat{m}$ with equal amplitude on the two
$(\hat{n} \cdot \overrightarrow{\sigma_2})$-eigenstates, and for a sufficiently weak interaction strength the action of the operator, however, is equivalent to a rotation towards one of the eigenstates, and hence the similar expressions (\ref{eq:real1},\ref{eq:imag1}). One consequence of the non-unitarity of the operator $e^{g Im \left\langle A \right\rangle _w
\cdot (\hat{n} \cdot \overrightarrow{\sigma_2})}$ is that any qubit state with its Bloch vector direction pointing towards the equator of the Bloch sphere is being bent towards the same eigenstate. Interestingly, for larger spins, the non-unitary operator $e^{\theta \hat{n}\cdot \vec{S}}$, is known to generate minimum uncertainty states for the spin components perpendicular to $\hat{n}$ \cite{Agarwal1990}, which may, in the limit of very large spins reflect the connection of weak values with the variances of the free-particle meter degrees of freedom pointed out by Jozsa \cite{Jozsa2007}.

Let us point out that the whole experimental set-up, so far described in terms of a meter system extracting the weak value of a system observable, may just as well be seen from the perspective of the meter qubit, interacting with an ancilla system, which is measured (the post-selection), and thus affecting a generalized measurement on the meter-system. The operator,
$e^{- i g \left\langle A \right\rangle _w
\cdot (\hat{n} \cdot \overrightarrow{\sigma_2})}$, acting on the meter, due to this measurement is thus
a measurement operator in a general POVM measurement,
and it is natural that it contains both unitary and non-unitary elements. In this perspective, it is also interesting to point out, that if the meter system is itself entangled with a third particle, a unitary operation acting only on the meter will not influence their mutual entanglement, while the non-unitary operation described above exactly has this capacity, as it may increase and reduce the state $|0\rangle$ and $|1\rangle$  amplitudes in a state of the form $a|00\rangle + b|11\rangle$, and hence bring the state closer to the maximally entangled state. For a more general analysis of this possibility applied to optical fields, see \cite{Menzies2008}.

\section{Weak value measurements on qubit systems with qubit meters}

\subsection{Measurement on a single qubit}

We will now address the simple case of a single qubit prepared and post selected in pure quantum states, and coupled weakly to another qubit system, which allows the read-out of the weak value of any qubit observable.

\subsubsection{Pure pre- and post-selected states}

We use the Bloch vector representation and parametrize the pre- and post-selected qubit states by unit vectors $\hat{r}_i,\hat{r}_f$,
\bgeq
\ket{\psi_i}\bra{\psi_i} = \frac{1}{2} \left( I +  \hat{r}_i \cdot \overrightarrow{\sigma}\right),
\edeq
\bgeq
\ket{\psi_f}\bra{\psi_f} = \frac{1}{2} \left( I +  \hat{r}_f \cdot \overrightarrow{\sigma}\right).
\edeq
Any qubit operator can be expanded as a combination of the identity matrix and the three Pauli matrices, and since the identity operator leads to only trivial results, we restrict hereafter the analysis to observables that can be parametrized $A=\hat{n}\cdot \overrightarrow{\sigma}$.

For the qubit
pre-selected in state $\ket{\psi_i}$ and post-selected in state $\ket{\psi_f}$ we thus apply the definition (\ref{eq:weak}), \bgeq
\left\langle \hat{n}\cdot \overrightarrow{\sigma} \right\rangle _w = \frac{\bra{\psi_f} \hat{n}\cdot \overrightarrow{\sigma} \ket{\psi_i}}{\left\langle \psi_f | \psi_i \right\rangle}.
\edeq
A straightforward calculation yields
\bgeqn
\left\langle \hat{n}\cdot \overrightarrow{\sigma} \right\rangle _w &=& \frac{\bra{\psi_f}\hat{n}\cdot \overrightarrow{\sigma}\ket{\psi_i} \left\langle \psi_i | \psi_f \right\rangle }{\left|\left\langle \psi_f | \psi_i \right\rangle \right|^2} \nonumber \\
&=&\frac{Tr\{ (I +  \hat{r}_f \cdot \overrightarrow{\sigma}) \hat{n}\cdot \overrightarrow{\sigma} \left( I +  \hat{r}_i \cdot \overrightarrow{\sigma} \right) \}}{Tr\{ (I +  \hat{r}_f \cdot \overrightarrow{\sigma}) \left( I +  \hat{r}_i \cdot \overrightarrow{\sigma} \right) } \nonumber \\
&=& \hat{n}\cdot \overrightarrow{w}
\edeqn
where $\overrightarrow{w}$ is a complex vector in $\mathcal{C}^3$, given by
\bgeq
\overrightarrow{w} = \frac{\hat{r}_i + \hat{r}_f + i \left( \hat{r}_i \times \hat{r}_f \right)}{1+\hat{r}_i\cdot \hat{r}_f}.
\label{wvectordef}
\edeq
We define two real unit vectors $\hat{r}$ and $\hat{s}$  in $\mathcal{R}^3$ by
$\hat{r} = \frac{\hat{r}_i + \hat{r}_f}{\left| \hat{r}_i + \hat{r}_f\right|}$,
and $\hat{s} = \frac{\hat{r}_i \times \hat{r}_f}{\left| \hat{r}_i \times \hat{r}_f\right|}$,
and denote the angle between $\hat{r}_f$ and $\hat{r}_i$ by $2 \theta$,
then
\bgeq
\overrightarrow{w} = \frac{1}{\cos \theta} \hat{r} + i \cdot \tan \theta \cdot \hat{s}. \label{wvectortheta}
\edeq
With this representation, the real part and imaginary part of the weak value are given respectively by
\bgeqn
Re \left\langle \hat{n}\cdot \overrightarrow{\sigma} \right\rangle _w &=& \frac{1}{\cos \theta} \hat{n}\cdot \hat{r} \\
Im \left\langle \hat{n}\cdot \overrightarrow{\sigma} \right\rangle _w &=& \tan \theta \hat{n}\cdot \hat{s} .
\edeqn
When $\ket{\psi_f}$ and $\ket{\psi_i}$ become almost orthogonal, then $\hat{r}_f$ and $\hat{r}_i$ become almost anti-parallel, $\theta \rightarrow \frac{\pi}{2}$, and both the real part and imaginary part of the weak value diverge. The probability to post-select a quantum state with its Bloch vector almost anti-parallel to the initial state, however, becomes vanishingly small near this divergence.

The real and imaginary part of the weak value for a single qubit can approach infinity when the pre and post-selected states becomes almost orthogonal.  Using (\ref{wvectordef}) and (\ref{wvectortheta}) we can easily show, however, that
\bgeq
|\left\langle \psi_f | \psi_i \right\rangle| \cdot \overrightarrow{w} = \hat{r} + i\cdot \sin \theta \cdot \hat{s},
\edeq
which implies that the real part, the imaginary part and the absolute value of the weak value
$\left\langle \hat{n}\cdot \overrightarrow{\sigma} \right\rangle _w = \hat{n} \cdot \overrightarrow{w}$
are all bounded by $1/|\left\langle \psi_f | \psi_i \right\rangle|$.

\subsubsection{General pre- and post-selected states}

Suppose the pre-selected state of the qubit is a general state $\rho_i$, and the post-selection is represented by
a measurement operator $P_f$ ($\sum_f P_f = I$). We can then formally define a state by $\rho_f = P_f /Tr P_f$, and {\sl call} it the post-selected state.

The weak value of the observable $\Omega =(\hat{n} \cdot \overrightarrow{\sigma})$, accessible by the same procedure of coupling to a meter system as for pure states, is in the case of mixed state given by
\bgeqn
\left\langle \Omega \right\rangle _w &=&  \frac{Tr\{ \rho_f \Omega \rho_i \}}{Tr \{ \rho_f \rho_i \}}.
\edeqn

For a general quantum system and a general observable $\Omega$, if the pre- and post-selected states are pure states,
then it is straightforward to show that both the real part, the imaginary part and the absolute value of the weak value
of $\Omega$
are bounded by $|e_m|/|\left\langle \Psi_f | \Psi_i \right\rangle|=|e_m|/\sqrt{Tr\{ \rho_f \rho_i \}}$,
where $e_m$ is the eigenvalue of $\Omega$ with the maximum absolute value. If the pre- and post-selected states are mixed states, however, the absolute value of the weak value is
no longer bounded by $|e_m|/\sqrt{Tr\{ \rho_f \rho_i \}}$. But it is easy to show that the absolute value
of the weak value is then bounded by $|e_m|/Tr\{ \rho_f \rho_i \}$.

If the system is coupled to an environment in the time interval between the pre- and post-selection, further contributions to the environment induced decoherence have to be taken into account \cite{SHO08}, but we will not address this issue in the present paper.

\subsection{Measurement on two qubits}

It is for several reasons interesting to extend the analysis to the case where the quantum system is a pair of qubits. In particular, this extends the analysis of weak values to operator products and correlation functions, and with both the investigated system and the meter system being composed of qubits, we may interchangeably describe multiple measurements on a single qubit and single measurements on composite systems, depending on what we define as post-selection and measurements, respectively.

\subsubsection{Product pure states as pre-selected and post-selected states}

Suppose two qubits are pre-selected in the product pure state
$\ket{\Psi_i}= \ket{\psi_i^a} \otimes \ket{\psi_i^b}$,
\bgeqn
\ket{\psi_i^a}\bra{\psi_i^a} = \frac{1}{2} \left( I +  \hat{r}_i^a \cdot \overrightarrow{\sigma_a}\right), \\
\ket{\psi_i^b}\bra{\psi_i^b} = \frac{1}{2} \left( I +  \hat{r}_i^b \cdot \overrightarrow{\sigma_b}\right),
\edeqn
and post-selected in the product pure state $\ket{\Psi_f}= \ket{\psi_f^a} \otimes \ket{\psi_f^b}$,
\bgeqn
\ket{\psi_f^a}\bra{\psi_f^a} = \frac{1}{2} \left( I +  \hat{r}_f^a \cdot \overrightarrow{\sigma_a}\right), \\
\ket{\psi_f^b}\bra{\psi_f^b} = \frac{1}{2} \left( I +  \hat{r}_f^b \cdot \overrightarrow{\sigma_b}\right).
\edeqn
The weak value of a product observable $\Omega =(\hat{n}^a \cdot \overrightarrow{\sigma_a})\otimes(\hat{n}^b \cdot \overrightarrow{\sigma_b})$ on the two qubits is given by the usual formalism
\bgeq
\left\langle \Omega \right\rangle _w = \frac{\bra{\Psi_f}\Omega \ket{\Psi_i}}{\left\langle \Psi_f | \Psi_i \right\rangle}.
\edeq
It is straightforward to show that
\bgeqn
\left\langle \Omega \right\rangle _w &=& \left( \hat{n}^a \cdot \overrightarrow{w_a}\right) \left( \hat{n}^b \cdot \overrightarrow{w_b}\right)\nonumber  \\
&=& \left\langle \hat{n}^a \cdot \overrightarrow{\sigma_a} \right\rangle _w  \left\langle \hat{n}^b \cdot \overrightarrow{\sigma_b} \right\rangle _w , \label{weakvalueofproduct}
\label{eq:factoring}
\edeqn
where $\overrightarrow{w_\alpha}$ ($\alpha =a,b$) is defined similarly as in (\ref{wvectordef}) with
$\hat{r}_i$ and $\hat{r}_f$ replaced by $\hat{r}_i^\alpha$ and $\hat{r}_f^\alpha$ respectively.
Eq. (\ref{weakvalueofproduct}) shows that for pre- and post- selected pure product states, the weak value of
a product operator is simply the product of the individual weak values.

An experiment measuring a single physical quantity, provides a  single real outcome,
and as we learned in the previous section, different experiments are thus needed to get access
to both the real and imaginary parts of the weak value of an observable. The real part of the weak value
$\langle \Omega \rangle_w$, may thus be obtained in one type of experiment while the imaginary part is available in another. Using the factorization property (\ref{eq:factoring}), we learn that
\bgeq
Re\langle \Omega \rangle_w = Re(\hat{n}^a \cdot \overrightarrow{w_a})
Re(\hat{n}^b \cdot \overrightarrow{w_b}) - Im (\hat{n}^a \cdot \overrightarrow{w_a})  Im (  \hat{n}^b \cdot \overrightarrow{w_b}),
\edeq
which interestingly implies that to replace the single measurement giving access to the real part of the weak value of a product operator by separate measurements on the two qubits, one must perform four measurements to obtain both the real and imaginary parts of the individual weak values of either qubit.

\subsubsection{Mixed pre- and post-selected states of two qubits}

General mixed pre- and post selected states $\rho_i$ and $\rho_f$ can be written as
\bgeqn
\rho_i &=& \frac{1}{4} \left( I\otimes I + \overrightarrow{r_i^a} \cdot \overrightarrow{\sigma} \otimes I
+ I \otimes \overrightarrow{r_i^b} \cdot \overrightarrow{\sigma} + \sum_{m,n=1}^{3} \omega_{mn}^{i} \sigma_m \otimes \sigma_n \right) , \\
\rho_f &=& \frac{1}{4} \left( I\otimes I + \overrightarrow{r_f^a} \cdot \overrightarrow{\sigma} \otimes I
+ I \otimes \overrightarrow{r_f^b} \cdot \overrightarrow{\sigma} + \sum_{m,n=1}^{3} \omega_{mn}^{f} \sigma_m \otimes \sigma_n \right) ,
\edeqn
and the weak value of a product observable $\Omega =(\hat{n}^a \cdot \overrightarrow{\sigma_a})\otimes(\hat{n}^b \cdot \overrightarrow{\sigma_b})$
can be shown to be
\bgeqn
\left\langle \Omega \right\rangle _w &=&  \frac{Tr\{ \rho_f \Omega \rho_i \}}{Tr \{ \rho_f \rho_i \}}  \\
&=&
\frac{1}{1+\overrightarrow{r_f^a}\cdot \overrightarrow{r_i^a}+ \overrightarrow{r_f^b}\cdot \overrightarrow{r_i^b} +\sum_{mn} \omega_{mn}^f \omega_{mn}^i} \left\{
\sum_{\alpha \beta} n_{\alpha}^a (\omega_{\alpha \beta}^f + \omega_{\alpha \beta}^i + r_{f\alpha}^a r_{i\beta}^b + r_{i\alpha}^a r_{f\beta}^b ) n_{\beta}^b
\right. \nonumber \\
& & -\sum_{\alpha \beta} \sum_{mnm'n'} n_{\alpha}^a n_{\beta}^b \varepsilon_{\alpha m m'} \varepsilon_{\beta n n'} \omega_{mn}^i \omega_{m' n'}^f
\nonumber \\
& & \left. +i \left[
\sum_{\alpha \beta} \sum_{m \gamma} n_{\alpha}^a n_{\beta}^b \varepsilon_{\alpha \gamma m} (\omega_{m \beta}^f r_{i\gamma}^a - \omega_{m\beta}^i r_{f \gamma}^a) +
\sum_{\alpha \beta} \sum_{n \gamma} n_{\alpha}^a n_{\beta}^b \varepsilon_{\beta \gamma n} (\omega_{\alpha n}^f r_{i\gamma}^b - \omega_{\alpha n}^i r_{f \gamma}^b).
\right]
\right\}
\edeqn
If the pre- and post-selected states are Werner states, namely,
$\overrightarrow{r_f^a}= \overrightarrow{r_i^a} = \overrightarrow{r_f^b}= \overrightarrow{r_i^b}=0 $
and $\omega_{mn}^f =\lambda_f \delta_{mn}$, $\omega_{mn}^i =\lambda_i \delta_{mn}$ ($-1\leq \lambda_{i(f)} \leq 1/3$), we have
\bgeq
\left\langle \Omega \right\rangle _w = \frac{1}{1+ 3 \lambda_f \lambda_i} (\lambda_f +\lambda_i) \hat{n}^a \cdot \hat{n}^b .
\label{wernerweakvalue}
\edeq
As for pure states we find divergent weak values in the case where the pre- and post-selected states belong to orthogonal subspaces (values $-1$ and $1/3$ for the $\lambda$-parameters correspond to density operators equal to the projection on the joint spin singlet entangled state and on its orthogonal complement). When traced over one qubit, the other qubit is in the fully mixed state with a vanishing mean effective spin, but the spins are correlated or anti-correlated in the Werner states, and this correlation is also found in the weak values with pre- and post-selected Werner states.

\section{Experimental implementation with trapped ions}

To carry out an experiment to illustrate weak value measurements on bipartite quantum systems, we need to prepare an initial state of a system and meter qubit, we must implement their mutual interaction $H_{int}$, and we must be able to post-select the final state of a system qubit, while detecting the change of the meter observable giving access to the weak value of interest. The ability to prepare and detect arbitrary single qubit states and to control the interaction between two qubits as in our $H_{int}$ is a major requisite for any attempt to construct a quantum computer. By definition, a universal quantum computer is able to implement any unitary operation on its register qubits, and hence provides a natural candidate for weak value measurements \cite{BDS08}. Furthermore, experimental proposals for quantum computers exist within a number of different systems, and while these proposals are difficult to scale to large numbers of qubits, several proposals are already sufficiently advanced to address few bits, and hence allow experimental verification of weak value predictions.

Very high fidelity operations with a very high read-out efficiency has been demonstrated in experiments with trapped ions, and we propose to demonstrate the above results on a pair of trapped ions. Trapped ions implement qubits as superposition states of two internal electronic states which can be optically pumped into one state with unit efficiency, and which can be driven into any superposition states by resonant laser-atom interaction. Read-out with as high as 99.99 \% fidelity \cite{Steane09} has been demonstrated using the shelving phenomenon in which one of the qubit states is repeatedly excited on a closed optical transition thus emitting a large number of photons if the ion populates the particular qubit states and none if it populates the other one. This read out scheme thus uniquely post selects the state of the system qubit, and it may also be used to perform a precise projective measurement of the electronic state of the meter qubit. Note that, if the post selection is not intended to produce one of the qubit electronic eigenstates, but some arbitrary state, one may either precede the eigenstate detection by the corresponding unitary rotation of the qubit, or one may apply coherent excitation from both eigenstate levels and thus identify "dark" and "bright" superposition states \cite{RoosKM}, which are distinguishable by the shelving mechanism.

What remains is the engineered interaction between the system and meter qubits in the form of the product of two projections of the effective spins of the two qubits. This interaction has been the cornerstone for multiple experiments on trapped ions in the past few years, and it can be implemented easily, for example by illuminating both ions with two different laser frequencies, which add up to twice the frequency needed to transfer a single ion between its qubit electronic states, but which are not resonant with the transitions in a single ion. If one takes the laser frequencies to be oppositely detuned from the ion frequencies by amounts slightly different from the common mode motional frequency of the ions in the trap, one derives a perturbative picture of either of the ions being virtually excited by the first photon together with a change of energy of the motion, followed by the absorption of the second photon, exciting the other ion and resonantly returning the common mode motion state to its initial state \cite{SM1999}. A wide choice of laser parameters allows elimination of the intermediate, virtually excited states, and leads precisely to an effective interaction of the desired form. Since our system is not subject to any other interaction terms, we do not need the interaction to be ultra short as in (6), but can take the time needed to accumulate the unitary interaction operator with a small value of $g$ in Eq.(7).

It is in fact also possible to apply laser parameters for which the elimination of states with motional excitation and a single ion excited is not justified, but for such parameters it is possible to adjust precisely the laser strengths and pulse durations, so that the undesired motional excitation is precisely removed at the end of the pulse, and again the effect of the laser-ion interaction leads to the desired unitary system-meter interaction with a strength $g$ that can be adjusted at will \cite{SM2000}. For an experimental verification of the latter, faster gates with finite $g$, and with extremely high fidelity, see for example \cite{Benhelm2008}. The two-bit interaction involves a product of specific spin-components $\hat{n}\cdot \overrightarrow{\sigma}$, which may indeed coincide with the ones ideal for the weak measurement experiments, and otherwise, the qubits can be appropriately rotated before and after the two-bit interaction between the system and meter is applied.

The weak value detection of a two-bit product operator needs the joint coupling of two system qubits to a third meter qubit. The latter is generally made possible by the universality of one- and two-bit gates in providing any arbitrary unitary evolution of the three-bit system, but in the case of trapped ions it may also be provided more directly by the three-bit effective interaction caused by a short of sequence of laser-atom interactions in the trap \cite{WSM}. In that paper it is for example shown how general products of three qubit operators can be implemented and how a single Hamiltonian involving differently detuned lasers provides a Toffoli- or Control-Control-NOT gate, which rotates the third qubit if both of the first qubits are in the qubit 1-state. This is readily formulated as a rotation conditioned on the product of the spin eigenstate projection operators for the two particles. Applying a small rather than a $\pi/2$ qubit rotation of the meter qubit, will thus enable the weak value measurement of the product operator.

The ability to carry out subsequent entanglement operations and multiple read out operations on a pair of trapped ions was used in a recent test of quantum contextuality with trapped ions \cite{context2009}, and we thus conclude that all ingredients are available for an experimental test of the measurability of the real and imaginary components of weak values for qubits and pairs of qubits with trapped ions.

\section{Conclusion}

In this paper, we have studied the properties of quantum weak values and possibilities to perform measurements that reveal their values. We have in particular addressed the anomalous aspects of these values being outside the spectral range of the operator in question and showing imaginary values, even for hermitian operators. The way that weak values are defined for systems in different pre- and post-selected states, and the way we formally interact with a meter systems to extract the weak values, clearly explain the possibility to reach very special results, and we have in this paper shown that by choosing a qubit meter system, we can use very similar procedures to obtain the real and imaginary parts of the weak value for arbitrary system operators. We imagine that studies of different meter systems may contribute further insight into the meaning of weak values and in particular on the duality between weak measurements on a system in pre- and post-selected states, and general quantum operations on the meter system caused by the interaction and post-section on the primary quantum system.

We have presented specific results for qubits and pairs of qubits in both pure and mixed states and derived methods to address these results experimentally with trapped ions. There is an ongoing discussion about the use of weak measurements for high precision purposes, and with the role played by ions and ion ensembles in atomic clocks, we anticipate that weak measurement strategies in these systems will be worth studying in more detail. Finally, fundamental tests of quantum mechanics based on the test of correlations and expectation values, suggest new investigations cf. the analysis of repeated measurements and simultaneous measurements of non-commuting observables by separate meter systems in \cite{BDS08}, the recent studies of the Leggett-Garg inequalities \cite{WJ08,GABLOWP09} and of quantum contextuality \cite{context2009}.

\section*{Acknowledgements}

S. W. wishes to acknowledge support from the NNSF of China (Grant No. 10604051), the CAS, and the National Fundamental Research Program.

\end{document}